\pdfoutput=1 
\documentclass{aa}
\usepackage[varg]{txfonts}
\usepackage{natbib,twoopt}
\usepackage{array}
\usepackage{graphicx}
\usepackage{gensymb}
\usepackage{amsmath}
\usepackage{subcaption}
\usepackage{float}

\usepackage{natbib,twoopt}
\usepackage[breaklinks=true]{hyperref} 
\bibpunct{(}{)}{;}{a}{}{,}             
\makeatletter
  \newcommandtwoopt{\citepads}[3][][]{\href{http://adsabs.harvard.edu/abs/#3}%
    {\def\hyper@linkstart##1##2{}%
     \let\hyper@linkend\@empty\citepalp[#1][#2]{#3}}}
  \newcommandtwoopt{\citeppads}[3][][]{\href{http://adsabs.harvard.edu/abs/#3}%
    {\def\hyper@linkstart##1##2{}%
     \let\hyper@linkend\@empty\citepp[#1][#2]{#3}}}
  \newcommandtwoopt{\citeptads}[3][][]{\href{http://adsabs.harvard.edu/abs/#3}%
    {\def\hyper@linkstart##1##2{}%
     \let\hyper@linkend\@empty\citept[#1][#2]{#3}}}
  \newcommandtwoopt{\citepyearads}[3][][]%
    {\href{http://adsabs.harvard.edu/abs/#3}
    {\def\hyper@linkstart##1##2{}%
     \let\hyper@linkend\@empty\citepyear[#1][#2]{#3}}}
\makeatother

\usepackage{breakurl}

\begin{document}

\title{Green Bank Telescope: Overview and analysis of metrology systems and pointing performance}

\author{E.~White\inst{1}
\and F.~D.~Ghigo\inst{2}
\and R.~M.~Prestage\inst{3,4}
\and D.~T.~Frayer\inst{2}
\and R.~J.~Maddalena\inst{2}
\and P.~T.~Wallace\inst{5}
\and J.~J.~Brandt\inst{2}
\and D.~Egan\inst{2}
\and J.~D.~Nelson \inst{2}
\and J.~Ray\inst{2}
}

\institute{Marshall University
\and Green Bank Observatory 
\and West Virginia University
\and Deceased
\and RAL Space
}

\abstract{With a 100m$\times$110m off-axis paraboloid dish, the Green Bank Telescope (GBT) is the largest fully steerable radio telescope on Earth. A major challenge facing large ground-based radio telescopes is achieving sufficient pointing accuracy for observing at high frequencies, up to 116 GHz in the case of the GBT. Accurate pointing requires the ability to blindly acquire source locations and perform ad hoc corrections determined by observing nearby calibrator sources in order to obtain a starting position accurate to within a small margin of error of the target's location. The required pointing accuracy is dependent upon the half-power beamwidth, and for the higher-frequency end of GBT observing, this means that pointing must be accurate to within a few arcsecond RMS. The GBT’s off-axis design is advantageous in that it eliminates blockage of the dish and reduces sidelobe interference, and there is no evidence that the resulting asymmetric structure adversely affects pointing accuracy. However, factors such as gravitational flexure, thermal deformation, azimuth track tilt and irregularity, and small misalignments and offset errors within the telescope’s structure cause pointing inaccuracies. A pointing model was developed for the GBT to correct for these effects. The model utilizes standard geometrical corrections along with metrology data from the GBT’s structural temperature sensors and data from measurements of the track levels. In this paper we provide a summary of the GBT's pointing model and associated corrections, as well as a discussion of relevant metrology systems and an analysis of its current nighttime pointing accuracy.

}


\titlerunning{Green Bank Telescope: Metrology Systems and Pointing Performance}

\maketitle 

\section{Introduction}
\begin{figure}[h]
\centering
\includegraphics[width=8.5cm]{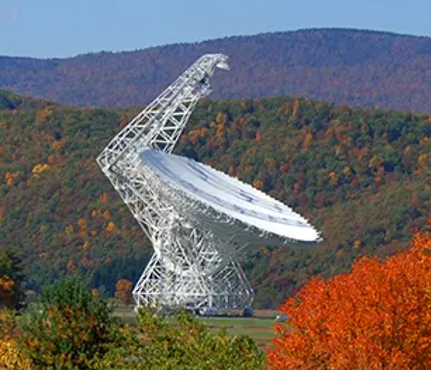}
\caption{Green Bank Telescope (GBO/Associated Universities, Inc./National Science Foundation).}
\label{fig:gbt1}
\end{figure}

The 100~meter Robert C. Byrd Green Bank Telescope (GBT; see Fig~\ref{fig:gbt1}) has an off-axis parabolic design with an unblocked aperture, which results in higher aperture efficiency and a cleaner diffraction pattern (which enables high-fidelity mapping), minimizes out-of-beam radio frequency interference (RFI), and avoids reflections that can compromise spectral baselines. The axis of the parabola is at the edge of the dish's surface. The large feed arm supports the receivers at the focus. Low-frequency receivers (0.29~GHz to 1.23~GHz) can be placed at the prime focus. The subreflector provides the secondary Gregorian focus, and up to eight receivers ride in a turret that can put any of them at the Gregorian focus. Thus, observing frequency bands can be changed in a few minutes. The suite of Gregorian receivers covers 1.15~GHz to 116~GHz. The aperture and beam efficiencies for the GBT at the high-frequency end of the observing band were measured recently; at 86~GHz, the aperture efficiency is $34.7\pm3.2\%$, and the main beam efficiency is $44.2\pm4.3\%$ \citep{frayer-etal-2019}. 

The moving structure is supported by a wheel and track system to rotate about the azimuth axis. Built at first to have good performance below 15~GHz, the adjustable surface and homologous design made it upgradable to function at higher frequencies, with a goal of acceptable performance up to 116~GHz, where the current nighttime aperture efficiency reaches approximately $20\%$ and the half-power beam width is about 6.4\arcsec \citep{frayer-etal-2019}. The subreflector is movable on six axes and is adjusted to compensate for flexure in the feed arm.  A 1~mm feature on the surface of the 32 m diameter track corresponds to a pointing offset of 6\arcsec, so the shape and flatness of the track must be measured to better than 0.5 mm accuracy.

Characterizing and correcting for these effects has been an ongoing project, and we are approaching reasonable performance at 116~GHz \citep{frayer2017, frayer2018, frayer-etal-2019}. The pointing model is the main focus of this paper, and while we will briefly address the GBT's focus tracking model, an in-depth treatment of this topic is left to other discussions.

Significant errors in pointing can reduce aperture efficiency and introduce uncertainty into radiometric calibrations of observations \citep{prestage2009green}. To minimize pointing errors, it is standard practice to implement pointing models for large single-dish telescopes. The 100~meter Effelsburg Telescope in Bonn, Germany \citep{wielebinski2011effelsberg}, the 50~meter Large Millimeter Telescope in Mexico \citep{schloerb2004large}, the 64~meter Sardinia Radio Telescope \citep{poppi2010highprecision}, the 30~meter Institute for Radio Astronomy in the Millimeter Range (IRAM) telescope \citep{penalver2000pointing, greve1996pointing}, and the 45~meter Nobeyama telescope in Japan \citep{ukita1999thermal} are among the large radio telescopes that have seen marked improvement in pointing performance after implementing a pointing model, achieving pointing accuracies down to a few arcseconds or less.

Despite the unique challenges posed by the GBT's off-axis design, the current pointing model achieves excellent performance and, when coupled with daily (or more frequent) adjustments to elevation and cross-elevation offset terms, enables the telescope to achieve offset nighttime pointing accuracy as good as 1.2\arcsec\  root mean squared (RMS) when a pointing calibration is done within 10$\degree$ of the source and within 1~hour of the observation. Without nearby pointing calibrations, the GBT can achieve 9\arcsec \ RMS pointing accuracy over the entire sky. These statements are detailed further in later sections. 

The GBT metrology system is made up of a suite of 19 structural temperature sensors mounted at various locations on the GBT structure, inclinometers placed on the elevation axle to measure distortions of the alidade structure, and 26-bit Heidenhain ROC 226 elevation and azimuth angle encoders (with a resolution of 0.02\arcsec each) that provide a readout of the GBT's indicated pointing location on the sky. The data products from these metrology systems are collected and incorporated into the GBT's pointing corrections in real time. In addition to the metrology data products, the pointing corrections also incorporate measurements of the azimuth track. 

The overall goal of this paper is to provide an overview of the development, performance, and upkeep of the GBT's pointing model, complete with a discussion of the metrology system's data products, how they contribute to the terms of the model, and what daily pointing calibrations are necessary to ensure optimum performance is achieved. In Sect. 2 we provide an overview of the challenges and factors involved in creating an accurate pointing model for the GBT as well as a summary of the terms included in the current pointing model and what physical effects they correct. In Sect. 3 the GBT's metrology systems and track model are described in detail. Section 4 contains a summary of GBT observing procedures and real-time pointing diagnostics and calibrations, as well as an overview of how the observational data and the data products from the metrology system are stored and later utilized. Section 5 presents a discussion of a comprehensive analysis that was executed using TPOINT modeling software -- which is utilized at several observatories around the world, including IRAM, Jodrell Bank, and the Atacama Large Millimeter/Submillimeter Array (ALMA) -- along with a summary of the current pointing model's overall performance at the GBT \citep{wallace1994tpoint}. Finally, we present our conclusions and suggestions for future work in Sect. 6.

\section{GBT pointing model}
\subsection{Overview of pointing-related systems}
The Precision Telescope Control System (PTCS) at the Green Bank Observatory (GBO) is responsible for the maintenance and improvement of GBT performance at high frequencies. The PTCS has implemented a variety of systems on the GBT that have contributed to improved observing quality. In addition to the pointing model, which we will discuss in more detail, the GBT also benefits from a focus tracking model, a finite element model (FEM), thermal Zernike corrections, and a gravity Zernike model to implement surface corrections; the pointing model itself depends on a combination of the refraction, gravity Zernike, and focus tracking models and on the FEM. The focus tracking model corrects for translations of the subreflector due to gravitational deflection of the feed arm as a function of elevation \citep{prestagebalser2005, ghigo2001focustracking}. Regular focus scans, conducted at the same time that pointing calibration measurements are done (see Sect. 4), are required in addition to the focus tracking model in order to achieve optimum focus. The focus tracking model has the functional form 

\begin{equation}
    F = a + b*cos(El) + c*sin(El)
.\end{equation}Here, El represents elevation, and a, b, and c are constants \citep{balser2006, prestagebalser2005}. These constant coefficients are checked and updated (if necessary) every few years; more frequent changes are not required because the parameters change so slowly. The telescope control software corrects for the effects of atmospheric refraction as a function of elevation before the pointing model is applied \citep{condon2004refraction, maddalena1994refraction, maddalena2002refraction}. 

Maintaining the accuracy of the GBT's primary surface is one requirement for optimum observing quality and pointing performance. Due to the GBT's offset feed design, the structure is not completely homologous, so additional corrections must be applied to adjust for nonhomologous gravitational deformations. The FEM is used to calculate corrections for the nonhomologous behavior of the dish, and a set of Zernike coefficients is used to model residual gravitational effects and thermal effects \citep{nikolic2007oof, hunter2009removal, hunter2011holography, maddalena2014}. These corrections, along with corrections for actuator misalignment (zero-point errors), are communicated to the GBT's active surface actuators to compensate for surface inaccuracies, providing significantly improved observational performance at high frequencies.  

It should be noted that different pointing models and focus tracking models are used when observing with the Gregorian focus receivers and the prime focus receivers. In this paper we discuss the models used for the Gregorian focus since this is where the high-frequency receivers are mounted and since pointing accuracy is essential at the higher end of the GBT's frequency coverage due to the smaller beam sizes \citep{balser2001pffocustrack, balser2001pfpointing,  balser2001pfallsky}.  

\subsection{Sources of pointing error}
The GBT's pointing model is responsible for correcting several factors that degrade the telescope's ability to point exactly where it is commanded to. In outlining the factors that affect the GBT's pointing accuracy, we first consider sources of pointing error that are universal for large altazimuth telescopes. The most basic sources of error, which affect nearly all large telescopes, include gravitationally induced vertical flexure, misalignment of the roll axis (i.e., tilt of the azimuth track for the GBT), azimuth and elevation encoder errors, and deviations from perpendicularity between the telescope's pointing axis and its elevation axis as well as between the azimuth axis and the elevation axis  \citep{wallace2008concise}. These error sources -- which are represented as terms in the current GBT model -- form the central framework of the pointing model.

The GBT's size and imbalanced design, along with its highly constrained alidade structure, make it susceptible to flexure resulting from a variety of physical causes, some of which can be modeled and corrected for \citep{prestage2009green}. Thermal gradients across the GBT's form result in structural expansion and contraction, which can significantly degrade pointing performance \citep{constantikes2008gbt, condon2004refraction}. Specifically, thermal gradients can cause the GBT's backup structure (the complex network of beams that supports the dish) to shift, the feed arm to bend and distort, and the alidade structure to deform. Thermally induced distortion of the backup structure and feed arm can cause deflections of the GBT's pointing direction in both azimuth and elevation \citep{constantikes2007therm}.

The impact of the GBT's azimuth track was alluded to earlier with reference to roll-axis misalignment errors. In addition to the track's overall tilt with respect to some reference level, the track also has minute deviations on its surface that can cause torques on the pintle bearing and alidade, resulting in pointing errors of less than 1\arcsec \citep{constantikes2008gbt}.

The pointing model does not currently account for wind-induced error. Although wind can set up vibrational modes in the telescope's structure that result in degraded pointing performance, the dynamical nature of wind's impact on the GBT and the difficulty of modeling this has made it challenging to include a wind term in the current pointing model \citep{constantikes2008gbt}. When wind speed exceeds 16 m/s sustained over a 1-minute period, telescope motion is stopped, and once it reaches 18 m/s, the telescope is stowed. Beyond these extreme thresholds, however, it is recommended that for each frequency range certain threshold wind speeds are not exceeded; these threshold speeds decrease as observing frequency increases.\footnote{GBT Observer's Guide: \url{https://www.gb.nrao.edu/scienceDocs/GBTog.pdf}} For example, at X-band frequencies (around 10~GHz), it is recommended that wind speeds not exceed 13.5 m/s, and for the highest-frequency range covered by the GBT (around 115~GHz), average wind speeds should be below 3.5 m/s (at night) in order to achieve optimal performance \citep{maddalena-frayer2014, condon-balser2015}.

\subsection{The pointing model}
The pointing model is a set of equations that relate the telescope encoder coordinates to the observed place of a celestial object. The direction of the incoming radiation is calculated from the catalog position of an object in mean J2000.0 (or International Celestial Reference System (ICRS)) coordinates, converted to the telescope-centric azimuth and elevation coordinates by accounting for precession, nutation, aberration (annual and diurnal), and Earth rotation. Refraction is added to produce the observed coordinates: $AZ_{obs}$ and $EL_{obs}$.

The mount coordinates ($AZ_{mnt}$ and $EL_{mnt}$) refer to the telescope's encoder values.  Hence, the pointing model is represented by the difference between the two coordinate frames:\\
\vspace{-1em}

\begin{equation}
   \Delta X\_EL = \Delta(AZ)*cos(EL_{mnt}) = (AZ_{mnt} - AZ_{obs})*cos(EL_{mnt})
   \label{eq1}
\end{equation}
\vspace{-2em}
\begin{equation}
    \Delta EL =  EL_{mnt} - EL_{obs}
    \label{eq2}
,\end{equation}

\noindent where cross-elevation (X\_EL, or \(AZ*cos(EL)\)) is used instead of azimuth in the formulation of the model \citep{prestage2003analyzing, brandt2007}. Appendix~A includes a full formulation of the GBT's pointing model as it is currently applied in equation form.

The GBT pointing model consists of 23 terms that correct for the factors described in Sect. 2.2. Tables~\ref{table1} and \ref{auxtable} summarize these terms (Sect. 3 will include a more in-depth summary of the calculations of the terms in Table~\ref{auxtable}).

The ten terms listed in Table~\ref{table1} -- CAL, HSCE, HSSE HSSASE, HSCASE, IE, HESA, HECA, HESE, and HECE -- address the most standard correction factors outlined in Sect. 2.2 and depend only on the telescope's azimuth and elevation. It should be noted that the TPOINT term names encode the functional form of each term; for example, “HSSE” means a harmonic (H) term applied to cross-elevation (S) as a function of sinE (SE). The choice of functional forms follows the convention used in an early formulation of the model, and we have continued using that method. A disadvantage is that the azimuth axis tilt terms, having different coefficients for the cross-elevation and elevation effects, do not directly provide the actual tilt angles.

It is apparent from Table~\ref{table1} that the azimuth tilt terms and azimuth and elevation axis misalignment coefficients are quite small, on the order of a few arcseconds. This shows the excellent mechanical alignment achieved by the manufacturing and assembly procedures. On the other hand, the elevation terms HESE and HECE are quite large, due mostly to the large flexure of the feed arm that supports the receiver room and subreflector.

The 13 terms in Table~\ref{auxtable} are auxiliary terms, which depend on additional external information from the GBT's metrology systems. Section 3 provides a further discussion of these terms and the instrumentation systems that collect auxiliary metrology data for use in the pointing model.

\section{Metrology systems of the GBT}
\subsection{Temperature sensors}
A suite of 19 structural temperature sensors on the GBT record the temperatures at various locations on the telescope. The temperature sensors are accurate to within $\pm 0.3\degree$ C \citep{constantikes2003tempsensors} and send readings back to the antenna control software so that they can be incorporated into the GBT's pointing model in real time \citep{marganian2003}. These temperature readings are used to correct for pointing errors that result from structural deformation due to the uneven thermal expansion of the GBT. Table~\ref{table2} provides a summary of the temperature sensors' locations and labels, and Fig.~\ref{fig:temp1} shows the location of each temperature sensor. Ten of the auxiliary terms listed in the pointing model -- A2S, A3S, A4S, A5E, A6E, A7E, A8E, A9E, A10V, and A11A from Table~\ref{auxtable} -- are thermal terms, calculated from the readings of the structural temperature sensors. Table~\ref{table3} provides an in-depth summary of the calculations that determine the values for the ten thermal terms. These expressions estimate how the differences between temperatures at different locations on the structure cause deflections in elevation and cross-elevation.

Terms A2S, A3S, and A4S correct for cross-elevation errors that result from thermal deformation, as summarized in Tables~\ref{auxtable} and \ref{table3}, whereas A5E, A6E, A7E, A8E, and A9E correct for errors in elevation. A10V corrects for thermally induced non-perpendicularity between the elevation and azimuth axes, and A11A corrects for errors in azimuth.

\subsection{Inclinometers}
The GBT is also equipped with two dual-axis Wyler Zeromatic inclinometers that are mounted on the GBT’s elevation bearings, as shown in Fig.~\ref{fig:incl} \citep{constantikes2005aztrack}. The X components of the inclinometers (X1, X2) measure the angle of deflection around the elevation axis, whereas the Y components of the inclinometers (Y1, Y2) measure the tilt perpendicular to the elevation axis.

\subsection{Azimuth track model}

The inclinometers measure how much the GBT’s elevation axis tilts due to the surface unevenness of the azimuth track, which in turn affects the GBT’s pointing direction. Deflections due to temperature-induced effects, overall track tilt, and the alidade’s spring constant are removed from the raw measurements so that the inclinometers provide accurate readings of the elevation and cross-elevation tilt caused by track surface errors \citep{constantikes2005aztrack}. Usually twice a year, a series of measurements is made in which the GBT is rotated slowly in azimuth, and a continuous stream of X1, X2, Y1, and Y2 readings from the inclinometers is recorded and tabulated as a function of azimuth. Table~\ref{lamtable} summarizes the pointing corrections calculated from inclinometer readings.

A12E, A13V, and A14A are coefficients derived by fitting the pointing model in 2016. Their values are tabulated in the first three entries of Table~\ref{auxtable}.

\subsection{History of the azimuth track}
The azimuth track is supported by a concrete circular foundation that extends several meters down into bedrock of hard Devonian shale (the exact depth of the bedrock varies in different sections of the foundation). There are four reinforcing walls spaced 90$\degree$ from one another, extending from the center to the circular track support. Atop the concrete are the 48 steel base plates and atop those are the 48 wear plates on which the wheels ride. The moving structure is supported in four places by bogies of four wheels each (i.e., a total of 16 wheels).  In the case of bad weather with high winds, the four assemblies move to align with the four reinforcing walls for maximum stability.

The track wear plates began to deteriorate during the first few years of operation (after 2001) due to high wheel load – over $4.54x10^5$ kg ($10^6$ lbs) per wheel. Analysis of the track and foundation \citep{symmes2008improving} led to plans for the replacement of the track components \citep{anderson2008replacement}. In June through August of 2007 the track was replaced with higher strength base plates made of bridge steel and tougher and thicker wear plates with higher impact and fatigue strengths. The wear plates now have V-shaped joints instead of the previously used 45-degree miter joints, and they have the same width as the base plates beneath them. A layer of wear-resistant and lubricating material made of bronze, Teflon, and molybdenum is sandwiched between the wear plates and the base plates to reduce the chance of fretting wear. The plates are held together by tensioned studs that go all the way through both the wear plates and the base plates. The cement grout that fills the void between the base plates and the top of the concrete is a high strength, two-part epoxy grout that is poured in to completely fill the void.  During the installation of the new track, care was taken to make the track as level and flat as possible using laser trackers.

Every year during June to August the wear plates are lifted, inspected, and the bronze layer replaced. Every two years, the wear plates are shifted by  one-quarter track segment lengths. Plates that show excessive wear are replaced.  We generally take inclinometer measurements of the track (X1, X2, Y1, Y2) before and after the track inspections.

\begin{figure}[h]
\centering
\includegraphics[width=8cm]{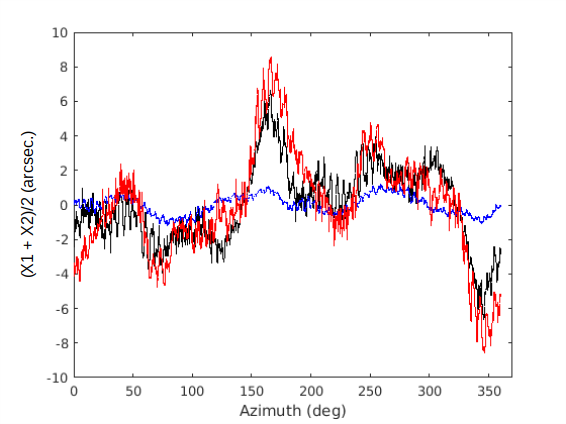}
\caption{Azimuth track measurements. The black trace was taken before the track replacement in April 2007, the blue trace was taken after the track replacement in September 2007, and the red trace was taken in September 2019.}
\label{recenttrackmap}
\end{figure}

The track measurement before and after the track replacement is plotted in Fig.~\ref{recenttrackmap}. The black line shows the track $(X1+X2)/2$ in April~2007 before the track replacement. One can see the deviation from flatness is about 6\arcsec. This corresponds to 1 mm on the track. The blue line is after the replacement, in which the maximum deviation is less than 1\arcsec. The red trace in Fig.~\ref{recenttrackmap} plots a recent track measurement taken in September 2019. Over the years since the track replacement, the deviations from flatness have gradually increased; at present, they are comparable to the state before the replacement. One may note that the deviations from flatness are near zero for azimuth 45º, 135º, 225º, and 315º. These locations are where the wheel assemblies are atop the reinforcing walls. Between these locations, the foundation may be gradually changing.

\begin{figure}[h]
\centering
\includegraphics[width=8cm]{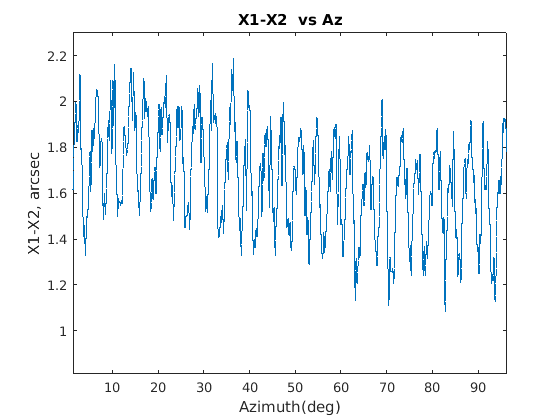}
\caption{Small-scale track deviations measured as the difference between X1 and X2 inclinometer readings.}
\label{smalltrackdev}
\end{figure}

Whereas Fig.~\ref{recenttrackmap} shows the large-scale deviations from flatness, there are also small-scale effects due to the wheels rolling from one track segment to the next. This can be seen in a plot of X1-X2 (see Fig.~\ref{smalltrackdev}). Each segment extends 7.5 degrees (1/48 of a circle).
This shows that the bumps rolling over a track segment amount to about 0.5\arcsec, or 83~$\mu$m.

\section{Observing and pointing calibration procedures}
\subsection{Implementation of the pointing model}
A pointing model is a parametric function that, given a source's catalog coordinates (corrected for factors such as the Earth's rotation, diurnal aberration, coordinate conversion, and refraction), outputs a set of pointing corrections that produce coordinates that are corrected for the sources of pointing error described in Sect. 1 \citep{prestage2003analyzing}. Mount coordinates (as described in Sect. 2.3) are the coordinates to which the telescope must be commanded by the telescope control software such that the target source is centered in the telescope's beam.

Applying the pointing model corrections (including thermal and azimuth track terms) along with local corrections obtained from pointing calibrations, standard receiver beam offset values, and tracking model corrections yields the total corrections in elevation and cross-elevation. These corrections must be sent to the servos so that they can adjust the telescope's position \citep{brandt2007, marganian2003}.

\subsection{Pointing calibration procedures}
\begin{figure}[h]
\centering
\includegraphics[width=9cm]{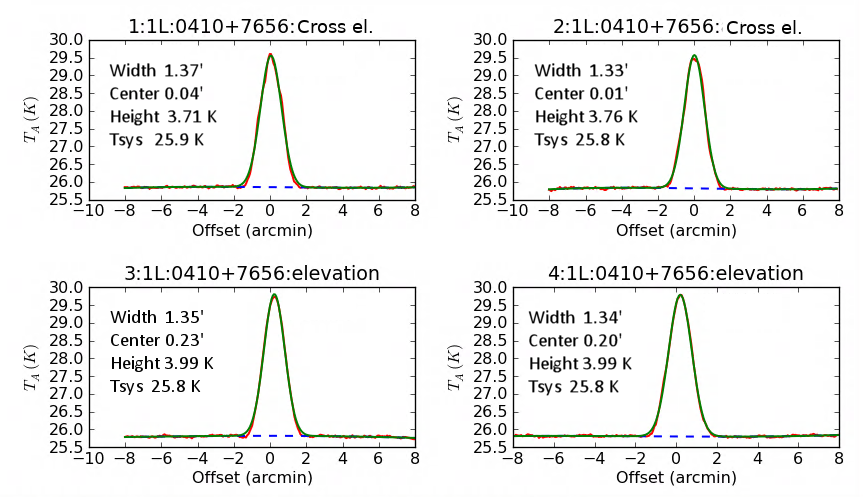}
\caption{Plot of the four component scans that make up a single pointing calibration measurement from a 2018 X-band pointing observation session; this diagram was generated using the GBT Fits Monitor (GFM) software. The dashed blue line represents the fitted baseline, the green line represents the modeled Gaussian, and the red line is real data. }
\label{gfm-plot}
\end{figure}

To achieve optimum pointing performance, high-frequency observers must perform pointing calibration procedures at the beginning of their observing session, as well as during the session as necessary (at high frequencies -- above about 40~GHz -- pointing calibrations should be performed every 30-60 minutes; at lower frequencies, pointing calibrations may only be needed every 1-3 hours or once at the start of the session).\footnote{GBT Observer's Guide: \url{https://www.gb.nrao.edu/scienceDocs/GBTog.pdf}} The pointing calibrations involve taking continuum observations of known, bright point sources in a pattern called a cross scan -- which consists of a series of forward and backward scans in the cross-elevation direction, as well as a forward scan and a backward scan in the elevation direction (see Fig.~\ref{gfm-plot}) \citep{heiles-maddalena1993, balser2003, condon2003quick, prestage2003analyzing}. Cross scans allow the observer to determine the pointing offsets in elevation and cross-elevation that remain after the pointing model has been applied, and these offsets can be input as local pointing corrections (LPCs) in the antenna control software. The LPCs are then added as offset terms in the pointing model by the control software, and this results in improved pointing performance. 

Our method of determining the values of coefficients for the GBT pointing model is to conduct one or more observing sessions in which we take several dozen pointing measurements of continuum point sources across the sky \citep{condon-yin2005}. We can extract pointing residuals from these observing sessions and use TPOINT modeling software to fit a model to the residuals, which will determine optimized values for the model's coefficients. Pointing coefficients are refit every few years, especially following any major changes to the track or other systems that may affect the model's overall performance.

\section{Pointing performance analysis}

Multiple times per year, observatory staff perform all-sky observing sessions that consist entirely of taking pointing and focus measurements of a selection of bright, unresolved radio sources (as these sessions are intended strictly to assess telescope performance, no science observations are interleaved with the pointing and focus measurements). Every few years, these data are used to create updated pointing models. These observing sessions provide a good sampling of data and can be used to characterize the GBT's pointing performance.

Data were selected based on a few requirements. All of the observations had to be taken with the X-band (8-10 GHz) receiver to ensure consistency; we note that the relative alignment between receivers is known to be stable to within about 0.5\arcsec or better, so pointing accuracy estimates at the X-band are also applicable at higher frequencies. In addition, only nighttime measurements were used in the analysis; this is representative of most high-frequency observing sessions, which are performed at night to avoid excessive deformation due to daytime thermal effects. Finally, we also ran the data through a quality-control program to reject damaged scans that had been corrupted by wind gusts or other abnormalities. Since all of the pointing measurements were of unresolved radio sources, each scan profile could be accurately modeled as a Gaussian with a shape determined by the frequency of the observation. If the observed profile's full width at half maximum (FWHM), height, and central position differed significantly from the modeled Gaussian or from the other scans in the pointing calibration sequence (i.e., by a more than 50\% difference between the measured and modeled value for the FWHM, a more than 50\% difference between the compared heights of two scans in the pointing calibration, and/or by a more than 300\% difference between the central positions of two scans), then that pointing measurement would be rejected under the assumption that there was something wrong with the data. Only one pointing measurement (0.1\% of the total number of measurements) was rejected based on these requirements from our 2017-2018 data set.

After selecting, preprocessing, and cleaning the data, the telescope pointing analysis program, TPOINT, was used to assess the GBT's pointing performance. A total of 14 pointing sessions (comprising 754 pointing measurements) spanning from January 2017 through December 2018 were included in the data set. This data set, along with the GBT's currently used pointing model (developed in 2016), was also loaded into TPOINT. One factor that had to be corrected to ensure an accurate characterization of the GBT's pointing performance was the fact that the elevation and cross-elevation constant offset terms (corresponding to CAL and IE in Table~\ref{table1}) had to be modified whenever subreflector hardware updates were made in order to compensate for changes to the software zero points that resulted from the hardware changes. The CAL and IE values for different time periods within our two-year data set are summarized in Table \ref{cal_ie_table}. 

\begin{table}[ht!]
\centering
\caption{Values of TPOINT cross-elevation and elevation constant offset terms (CAL and IE) over time.}
\begin{tabular}{ c c c }
\hline
\hline
\textbf{Time Period} & \textbf{CAL} & \textbf{IE} \\
\hline
Jan. 2017 -- Sept. 2017 & $-88.51$ & ~707.98\arcsec \\
Nov. 2017 -- Aug. 2018 & $-82.61$\arcsec & 696.53\arcsec \\
Dec. 2018 & $-106.51$ & 682.176\arcsec \\
\hline
\end{tabular}
\label{cal_ie_table}
\end{table}

The model was then applied in two different ways to the 754 pointing measurements; first, the model was applied with the CAL and IE terms fixed to measure the blind pointing performance. Second, TPOINT was allowed to find the best value for the CAL and IE offset terms for every one of the 14 observing sessions to determine the observing performance that would be attained if the observer performed a pointing calibration at the start of an all-sky observing session. This measurement yielded the all-sky relative pointing performance. TPOINT calculated the RMS of the pointing residuals for the 754 pointing measurements when the model was applied both ways: for blind pointing, RMS~=~9.0\arcsec, and for all-sky relative pointing, RMS~=~5.1\arcsec. We note that in both cases the ten most outlying measurements were masked within TPOINT to further improve data quality.

TPOINT can also generate pointing residual plots, the most instructive of which have been included in Appendix~B. The scatter plots in Fig.~\ref{fig:scatter-plots} show the error distribution with respect to the central beam of the GBT. Figure \ref{fig:rve-plots} shows plots of total pointing residual versus azimuth and elevation, respectively. We note that in the TPOINT output, the total pointing residual, dR, is the difference between the observed direction, [$Az_{obs}$,$El_{obs}$], and the corresponding encoder direction, [$Az_{el}$,~$El_{el}$]; the two vectors are expressed as:

\begin{align}
    &x_{obs} = cos(Az_{obs})*cos(El_{obs}) \nonumber \\
    &y_{obs} = sin(Az_{obs})*cos(El_{obs}) \nonumber \\
    &z_{obs} = sin(El_{obs}) \\ \nonumber \\
    &x_{el} = cos(Az_{el})*cos(El_{el}) \nonumber \\
    &y_{el} = sin(Az_{el})*cos(El_{el}) \nonumber \\
    &z_{el} = sin(El_{el}).
\end{align}

\noindent The differences between the components are given by Eq.~\ref{differences}:

\begin{align}
    &dx = x_{el} - x_{obs} \nonumber \\
    &dy = y_{el} - y_{obs} \nonumber \\
    &dz = z_{el} - z_{obs.}
    \label{differences}
\end{align}

\noindent So, the total error can be expressed as $dR = \sqrt{dx^2 + dy^2 + dz^2}$, which approximates to $dR = \sqrt{dS^2 + dE^2}$ to the first order, where dS is error in cross-elevation and dE is error in elevation.

The residuals measured as dR show no obvious azimuth- or elevation-dependent trends. This indicates that the current pointing model accounts for the elevation- and azimuth-dependent effects inherent to the GBT's structure (such as azimuth track tilt).

Another key metric on the performance of the telescope pointing model is the offset pointing uncertainty. In practice, astronomers will point on a bright source with a known position to derive LPCs and then slew to a nearby target position for science observations. The offset pointing uncertainty is defined as the positional error when targeting a position on the sky based on the LPCs measured from a pointing source nearby in position and in time. A good model permits the use of pointing sources that are farther away from the science target and provides consistent measurements over longer periods of time under stable conditions.

\begin{figure}[h]
\centering
\includegraphics[width=8cm]{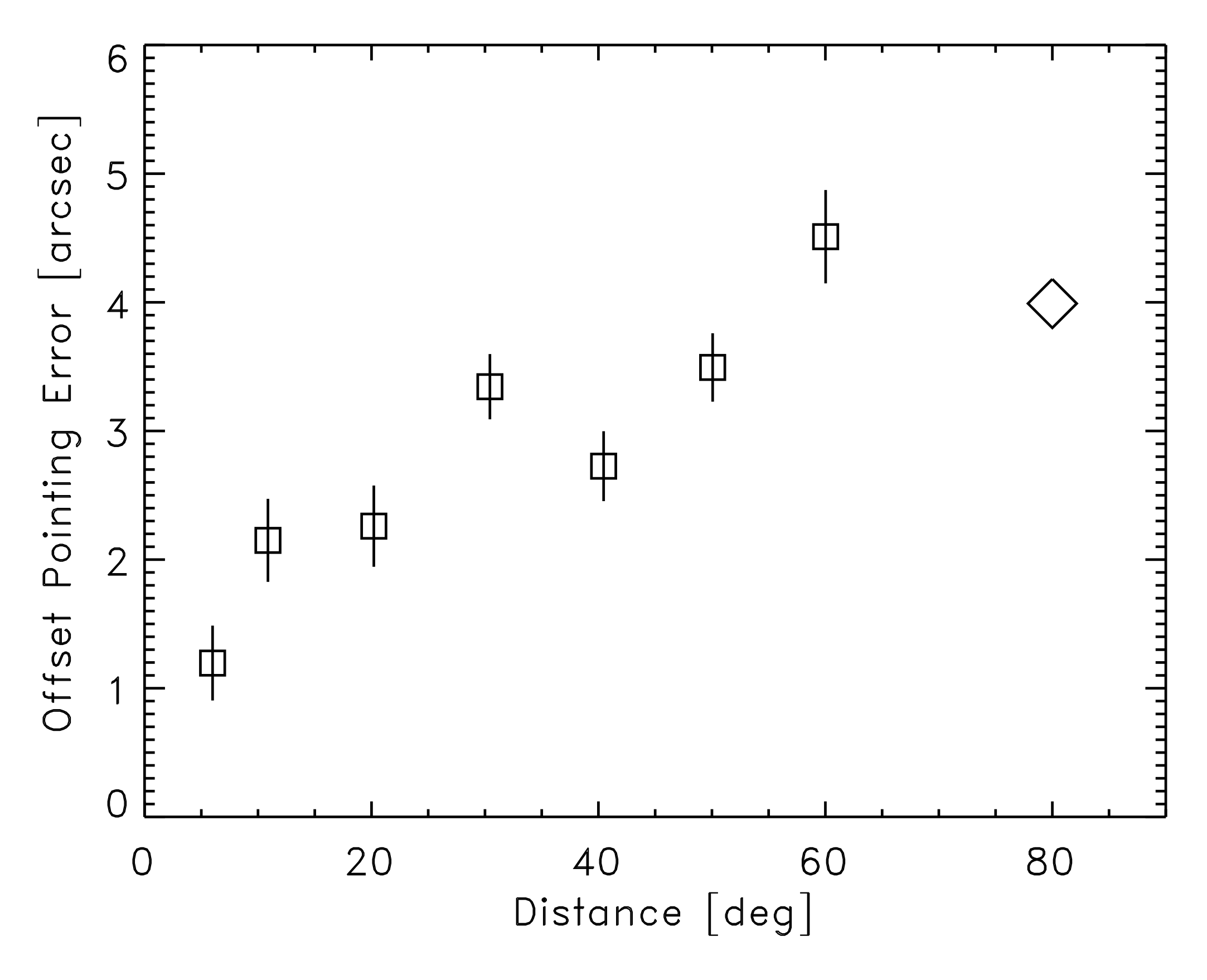}
\caption{Offset pointing error in arcseconds associated with the pointing model shown as squares as a function of distance from the pointing sources. The large diamond to the right shows the average error over all data (all-sky) taken over 7~hours.} 
\label{fig:offset_pnt}
\end{figure}

To quantify the offset pointing uncertainty, a long 7~hour X-band (9~GHz) pointing run was used that had a uniform all-sky distribution of 79 pointing measurements observed under stable nighttime conditions. The differences between the LPCs for a pair of measurements represent the offset pointing values. By computing the LPC differences for every possible combination of data pairs (3081), the offset pointing uncertainty was derived as a function of distance between the measurements. Figure~\ref{fig:offset_pnt} shows the results. The measurement errors need to be quantified to infer the errors due to the pointing model itself. The empirical measurement errors are computed by measuring the LPC differences from back-to-back measurements on the same source. The average radial error derived for an individual measurement ($\sigma_{obs}$) is 1.8\arcsec. The error due to the pointing model ($\sigma_{model}$) is

\begin{equation}
    \sigma_{model} = (\Delta xEl^2 + \Delta El^2 - (\sigma_{obs})^2)^{1/2}
,\end{equation}

\noindent where xEl represents cross-elevation and El represents elevation. For sources within 10$\degree$ and taken within 45~minutes of time, a model offset pointing error of 1.20~arcsec is derived. This error increases slightly with distance by computing values within 45~minutes of time for a range of distances. The average all-sky offset pointing error over the entire 7~hours is only 4~arcsec. An all-sky offset pointing uncertainty of 4~arcsec over an entire night suggests that users can reasonably point once per observing session for low-frequency observations with a large telescope beam (e.g., < 5 GHz), but for high-frequency observations (e.g., > 30 GHz) nearby pointing observations should be made regularly (more frequently than once per hour based on time variations). The overhead time required for a pointing and focus calibration is 4 minutes for the pointing measurement and 1 minute for the focus measurement, plus slew time at about 16 degrees per minute. \footnote{GBT Observer's Guide: \url{https://www.gb.nrao.edu/scienceDocs/GBTog.pdf}} 

\section{Conclusions}
We have demonstrated that the GBT is able to achieve very good blind pointing performance at high frequencies by utilizing a pointing model composed of standard geometric correction terms as well as thermal and track-dependent terms; the performance can be further improved by taking pointing calibrations at the beginning of an observing session to account for offset shifts. The GBT's metrology systems provide data products in real time that allow the model to correct for the thermal environment and track irregularities over the course of each observing session.

In the future, it would be ideal to improve the GBT's blind pointing performance in order to reduce the number of pointing calibrations required over the course of each night and thus reduce overhead time for observers. Areas for future work include more effectively incorporating wind and feed arm correction measurements from the quadrant detector installed on the GBT. TPOINT modeling experiments are currently under way to determine further improvements that could be made to the terms of the model.

\section{Acknowledgements}
The authors would like to extend special thanks to the late Kim~Constantikes (formerly of NRAO), to past and present GBO and NRAO staff for their substantial contributions to developing the GBT pointing model and control systems, and to the anonymous referee for their very constructive and helpful feedback. Ellie~White would like to thank her family for their support, and Dr.~Jon~Saken and the Marshall University Physics Department for allowing her to conduct this research as part of a series of independent study courses.\\

We dedicate this paper in fond memory of our friend and colleague Dr.~Richard~Prestage, who contributed tremendously to this project before his passing. His expertise and wisdom are deeply missed. \\

This research was made possible by NASA West Virginia Space Grant Consortium, NASA Agreement $\#$~80NSSC20M0055. The Green Bank Observatory is a facility of the National Science Foundation operated under cooperative agreement by Associated Universities, Inc. 


\begin{thebibliography}{40}
\expandafter\ifx\csname natexlab\endcsname\relax\def\natexlab#1{#1}\fi

\bibitem[{Anderson {et~al.}(2008)Anderson, Symmes, \&
  Egan}]{anderson2008replacement}
Anderson, R., Symmes, A., \& Egan, D. 2008, in Ground-based and Airborne
  Telescopes II, Vol. 7012, International Society for Optics and Photonics,
  701237

\bibitem[{Balser {et~al.}(2001{\natexlab{a}})Balser, Ghigo, Maddalena, \&
  Langston}]{balser2001pffocustrack}
Balser, D., Ghigo, F., Maddalena, R.~J., \& Langston, G. 2001{\natexlab{a}},
  GBT Commissioning Memo, 12

\bibitem[{Balser {et~al.}(2001{\natexlab{b}})Balser, Maddalena, Ghigo, \&
  Langston}]{balser2001pfpointing}
Balser, D., Maddalena, R.~J., Ghigo, F., \& Langston, G. 2001{\natexlab{b}},
  GBT Commissioning Memo, 5

\bibitem[{Balser {et~al.}(2001{\natexlab{c}})Balser, Maddalena, Ghigo, \&
  Langston}]{balser2001pfallsky}
Balser, D., Maddalena, R.~J., Ghigo, F., \& Langston, G. 2001{\natexlab{c}},
  GBT Commissioning Memo, 14

\bibitem[{Balser {et~al.}(2003)Balser, Condon, Constantikes, \&
  Prestage}]{balser2003}
Balser, D.~S., Condon, J.~J., Constantikes, K., \& Prestage, R.~M. 2003, NRAO
  Green Bank PTCS Project Note, 28

\bibitem[{Balser {et~al.}(2006)Balser, Nikolic, \& Prestage}]{balser2006}
Balser, D.~S., Nikolic, B., \& Prestage, R.~M. 2006, NRAO Green Bank PTCS
  Project Note, 49

\bibitem[{Brandt(2007)}]{brandt2007}
Brandt, J. 2007, NRAO Green Bank PTCS Project Note, 23

\bibitem[{Condon(2003)}]{condon2003quick}
Condon, J. 2003, NRAO Green Bank PTCS Project Note, 10

\bibitem[{Condon(2004)}]{condon2004refraction}
Condon, J. 2004, NRAO Green Bank PTCS Project Note, 35

\bibitem[{Condon \& Balser(2015)}]{condon-balser2015}
Condon, J. \& Balser, D.~F. 2015, GBT Dynamic Scheduling System Memo, 5

\bibitem[{Condon \& Yin(2005)}]{condon-yin2005}
Condon, J. \& Yin, Q. 2005, NRAO Green Bank PTCS Project Note, 36

\bibitem[{Constantikes(2003)}]{constantikes2003tempsensors}
Constantikes, K. 2003, NRAO Green Bank PTCS Project Note, 12

\bibitem[{Constantikes(2004)}]{constantikes2008gbt}
Constantikes, K. 2004, NRAO Green Bank PTCS Project Note, 53

\bibitem[{Constantikes(2005)}]{constantikes2005aztrack}
Constantikes, K. 2005, NRAO Green Bank PTCS Project Note, 44

\bibitem[{Constantikes(2007)}]{constantikes2007therm}
Constantikes, K. 2007, NRAO Green Bank PTCS Project Note, 25

\bibitem[{Frayer(2017)}]{frayer2017}
Frayer, D. 2017, NRAO GBT Project Memo, 296

\bibitem[{Frayer {et~al.}(2018)Frayer, Ghigo, \& Maddalena}]{frayer2018}
Frayer, D., Ghigo, F., \& Maddalena, R. 2018, NRAO GBT Project Memo, 301

\bibitem[{Frayer {et~al.}(2019)Frayer, Maddalena, White, Watts, Kepley, Li, \&
  Harris}]{frayer-etal-2019}
Frayer, D., Maddalena, R., White, S., {et~al.} 2019, NRAO GBT Project Memo, 302

\bibitem[{Ghigo {et~al.}(2001)Ghigo, Maddalena, Balser, \&
  Langston}]{ghigo2001focustracking}
Ghigo, F., Maddalena, R.~J., Balser, D., \& Langston, G. 2001, GBT
  Commissioning Memo., 7

\bibitem[{Greve {et~al.}(1996)Greve, Panis, \& Thum}]{greve1996pointing}
Greve, A., Panis, J.-F., \& Thum, C. 1996, Astronomy and Astrophysics
  Supplement Series, 115, 379

\bibitem[{Heiles \& Maddalena(1993)}]{heiles-maddalena1993}
Heiles, C. \& Maddalena, R. 1993, NRAO GBT Project Memo, 105

\bibitem[{{Hunter} {et~al.}(2009){Hunter}, {Mello}, {Nikolic}, {Mason},
  {Schwab}, {Ghigo}, \& {Dicker}}]{hunter2009removal}
{Hunter}, T., {Mello}, M., {Nikolic}, B., {et~al.} 2009, in 2009 USNC/URSI
  Annual Meeting, 1

\bibitem[{Hunter {et~al.}(2011)Hunter, Schwab, White, Ford, Ghigo, Maddalena,
  Mason, Nelson, Prestage, Ray, Ries, Simon, Srikanth, \&
  Whiteis}]{hunter2011holography}
Hunter, T., Schwab, F., White, S., {et~al.} 2011, Publications of The
  Astronomical Society of The Pacific - PUBL ASTRON SOC PAC, 123

\bibitem[{Maddalena \& Frayer(2014)}]{maddalena-frayer2014}
Maddalena, R. \& Frayer, D. 2014, GBT Dynamic Scheduling System Memo, 18

\bibitem[{Maddalena {et~al.}(2014)Maddalena, Frayer, Lashley-Colhirst, \&
  Norris}]{maddalena2014}
Maddalena, R., Frayer, D., Lashley-Colhirst, N., \& Norris, T. 2014, NRAO Green
  Bank PTCS Project Note, 76

\bibitem[{Maddalena(1994)}]{maddalena1994refraction}
Maddalena, R.~J. 1994, GBT Technical Memo.

\bibitem[{Maddalena {et~al.}(2002)Maddalena, Ghigo, Balser, \&
  Langston}]{maddalena2002refraction}
Maddalena, R.~J., Ghigo, F., Balser, D., \& Langston, G. 2002, GBT
  Commissioning Memo., 16

\bibitem[{Marganian(2004)}]{marganian2003}
Marganian, P. 2004, NRAO Green Bank PTCS Project Note, 17

\bibitem[{Nikolic {et~al.}(2007)Nikolic, Prestage, Balser, Chandler, \&
  Hills}]{nikolic2007oof}
Nikolic, B., Prestage, R., Balser, D., Chandler, C., \& Hills, R. 2007,
  Astronomy \& Astrophysics, 465, 685

\bibitem[{Penalver {et~al.}(2000)Penalver, Lisenfeld, \&
  Mauersberger}]{penalver2000pointing}
Penalver, J., Lisenfeld, U., \& Mauersberger, R. 2000, in Radio Telescopes,
  Vol. 4015, International Society for Optics and Photonics, 632--640

\bibitem[{Poppi {et~al.}(2010)Poppi, Pernechele, Pisanu, \&
  Morsiani}]{poppi2010highprecision}
Poppi, S., Pernechele, C., Pisanu, T., \& Morsiani, M. 2010, 7733

\bibitem[{Prestage \& Balser(2003)}]{prestage2003analyzing}
Prestage, R. \& Balser, D. 2003, NRAO Green Bank PTCS Project Note, 15

\bibitem[{Prestage \& Balser(2005)}]{prestagebalser2005}
Prestage, R. \& Balser, D. 2005, NRAO Green Bank PTCS Project Note, 42

\bibitem[{Prestage {et~al.}(2009)Prestage, Constantikes, Hunter, King, Lacasse,
  Lockman, \& Norrod}]{prestage2009green}
Prestage, R.~M., Constantikes, K.~T., Hunter, T.~R., {et~al.} 2009, Proceedings
  of the IEEE, 97, 1382

\bibitem[{Schloerb \& Carrasco(2004)}]{schloerb2004large}
Schloerb, F.~P. \& Carrasco, L. 2004, in Ground-based Telescopes, Vol. 5489,
  International Society for Optics and Photonics, 754--762

\bibitem[{Symmes {et~al.}(2008)Symmes, Anderson, \& Egan}]{symmes2008improving}
Symmes, A., Anderson, R., \& Egan, D. 2008, in Ground-based and Airborne
  Telescopes II, Vol. 7012, International Society for Optics and Photonics,
  701238

\bibitem[{Ukita(1999)}]{ukita1999thermal}
Ukita, N. 1999, Publications of the National Astronomical Observatory of Japan

\bibitem[{Wallace(1994)}]{wallace1994tpoint}
Wallace, P. 1994, Starlink User Note, 100

\bibitem[{Wallace(2008)}]{wallace2008concise}
Wallace, P.~T. 2008, in Advanced Software and Control for Astronomy II, Vol.
  7019, International Society for Optics and Photonics, 701908

\bibitem[{Wielebinski {et~al.}(2011)Wielebinski, Junkes, \&
  Grahl}]{wielebinski2011effelsberg}
Wielebinski, R., Junkes, N., \& Grahl, B.~H. 2011, Journal of Astronomical
  History and Heritage, 14, 3

\end{thebibliography}

\onecolumn
\begin{table}[ht!]
\caption{Geometric and gravity-related coefficients in the GBT pointing model.}  
\label{table1}
\centering
\begin{tabular}{ >{\centering\arraybackslash}p{5em} c c c >{\centering\arraybackslash}p{4em} >{\centering\arraybackslash}p{4em} }     
\hline\hline
\textbf{Coefficient Name} & \textbf{Coordinate} & \textbf{Functional Form} & \textbf{Description} & \textbf{Value (\arcsec)} & \textbf{Error (\arcsec)} \\
\hline
   CAL & $\Delta$AZcos(EL) & constant & Cross-elevation offset & -88.51 & 4.43 \\
   HSCE & $\Delta$AZcos(EL) & cos(EL) & Azimuth encoder offset & 39.60 & 3.11 \\
   HSSE & $\Delta$AZcos(EL) & sin(EL) & Elevation axle collimation & 2.55 & 3.55 \\
   HSSASE & $\Delta$AZcos(EL) & sin(AZ)*sin(EL) & North-South tilt of azimuth axis & -2.79 & 0.63 \\
   HSCASE & $\Delta$AZcos(EL) & cos(AZ)*sin(EL) & East-West tilt of azimuth axis & 1.83 & 0.44 \\
   IE & $\Delta$EL & constant & Elevation encoder offset & 707.98 & 4.76 \\
   HESA & $\Delta$EL & sin(AZ) & East-West tilt of azimuth axis & -1.86 & 0.30 \\
   HECA & $\Delta$EL & cos(AZ) & North-South tilt of azimuth axis & -4.61 & 0.36 \\
   HESE & $\Delta$EL & sin(EL) & Asymmetric gravity term & -647.81 & 3.50 \\
   HECE & $\Delta$EL & cos(EL) & Symmetric gravity term & -775.70 & 3.04 \\
\hline
\end{tabular}
\tablefoot{This table summarizes the geometric and gravity-related coefficients in the GBT pointing model and their corresponding functional forms, values (in arcseconds), and the error associated with the determination of each coefficient. Here, EL~stands for~mount~elevation and  AZ~for~mount~azimuth. The coefficients' signs follow TPOINT conventions, so in order to reproduce this model in a TPOINT session, these terms must be plugged in exactly as they appear in the table.}
\end{table}

\begin{table}[ht!]
\caption{Auxiliary coefficients in the GBT pointing model.} 
\centering
\begin{tabular}{ >{\centering\arraybackslash}p{5em} c c c >{\centering\arraybackslash}p{4em} >{\centering\arraybackslash}p{4em} }
\hline
\hline
\textbf{Coefficient Name} & \textbf{Coordinate} & \textbf{Functional Form} & \textbf{Description}  & \textbf{Value (\arcsec)} & \textbf{Error (\arcsec)} \\
\hline
A12E & $\Delta$EL & $\lambda_1$ & Track term & 0.705 & 0.15 \\
A13V & $\Delta$AZcos(EL) & $\lambda_2$ * sin(EL) & Track term  & -1.01 & 0.21 \\
A14A & $\Delta$AZcos(EL) & $\lambda_3$ * cos(EL) & Track term & -0.83 & 0.46 \\
A2S & $\Delta$AZcos(EL) & $\tau_2$ & Horizontal feedarm and el. bearings (thermal) & 6.79 & 1.57 \\
A3S & $\Delta$AZcos(EL) & $\tau_3$ & Backup structure and el. bearings (thermal) & -0.49 & 0.75 \\
A4S & $\Delta$AZcos(EL) & $\tau_4$ & Vertical feedarm (thermal) & 0.34 & 1.67 \\
A5E & $\Delta$EL & $\tau_5$ & Backup structure (thermal) & -1.65 & 1.94 \\
A6E & $\Delta$EL & $\tau_6$ & Horizontal feedarm and el. bearings (thermal) & -0.50 & 0.13 \\
A7E & $\Delta$EL & $\tau_7$ & Vertical feedarm (thermal) & -3.28 & 0.38 \\
A8E & $\Delta$EL & $\tau_8$ & Alidade (thermal) & 7.19 & 1.69 \\
A9E & $\Delta$EL & $\tau_9$ & Structural temp. average (thermal) & -0.35 & 0.09 \\
A10V & $\Delta$AZcos(EL) & $\tau_{10}$*sin(EL) & Alidade and el. bearings (thermal) & 10.73 & 4.24 \\
A11A & $\Delta$AZcos(EL) & $\tau_{11}$*cos(EL) & Alidade thermal term & -0.43 & 0.56 \\
\hline
\end{tabular}
\label{auxtable}
\tablefoot{This table summarizes the auxiliary coefficients in the GBT pointing model and their corresponding functional forms, values (in arcseconds), and the error associated with the determination of each coefficient. As in Table \ref{table1}, EL~stands for~mount~elevation and  AZ~for~mount~azimuth, and the coefficients' signs follow TPOINT conventions; therefore, in order to reproduce this model in a TPOINT session, these terms must be plugged in exactly as they appear above.}
\end{table}

\begin{table}[h]
\caption{GBT temperature sensors; see Fig. \ref{fig:temp1}.}
\centering
\begin{tabular}{ c c c }
\hline
\hline
\textbf{Sensor \#} & \textbf{Label} & \textbf{Location of sensor} \\
\hline
1 & TA1 & Alidade: right front \\
2 & TA2 & Alidade: left front \\
3 & TA3 & Alidade: right rear \\
4 & TA4 & Alidade: left rear \\
5 & TE1 & Elevation bearing: right \\
6 & TE2 & Elevation bearing: left \\
7 & TH2 & Horizontal feedarm: right \\
8 & TH1 & Horizontal feedarm: left \\
9 & TB1 & Backup structure vertex: right \\
10 & TB4 & Backup structure: center right \\
11 & TB3 & Backup structure: center left \\
12 & TB2 & Backup structure vertex: left \\
13 & TB5 & Backup structure: center \\
14 & TF2 & Vertical feedarm: right front \\
15 & TF3 & Vertical feedarm: right rear \\
16 & TF5 & Vertical feedarm: left rear \\
17 & TF4 & Vertical feedarm: left front \\
18 & TF1 & Feedarm tip \\
19 & TSR & Subreflector \\
\hline
\end{tabular}
\label{table2}
\tablebib{
\cite{constantikes2007therm}
}
\end{table}

\begin{table}[ht!]
\centering
\caption{Temperature correction calculations for each term.}
\begin{tabular}{ c c c }
\hline
\hline
\textbf{Label} & \textbf{Associated coefficient} & \textbf{Thermal correction} \\
\hline
$\tau_2$ & A2S & $(TH2-TE1)/2.0 - (TH1-TE2)/2.0$ \\
$\tau_3$ & A3S & $(TB1+TB4-TB2-TB3)/2.0 - TE1 + TE2$ \\
$\tau_4$ & A4S & $(TF2+TF3)/2.0 - (TF4+TF5)/2.0$ \\
$\tau_5$ & A5E & $(TB3+TB4+TB5)/3.0 - (TB1+TB2)/2.0$ \\
$\tau_6$ & A6E & $(TH2-TE1)/2.0 + (TH1-TE2)/2.0$ \\
$\tau_7$ & A7E & $(TF2+TF4)/2.0 - (TF3+TF5)/2.0$ \\
$\tau_8$ & A8E & $(TA1+TA2)/2.0 - (TA3+TA4)/2.0$ \\
$\tau_9$ & A9E & Average structural temperature \\
$\tau_{10}$ & A10V & $(TA2+TA3+TE2-TA1-TA4-TE1)/3.0$ \\
$\tau_{11}$ & A11A & $TA4-TA1-TA3+TA2$ \\
\hline
\end{tabular}
\label{table3}
\tablebib{
\cite{constantikes2007therm}
}
\end{table}

\begin{table}[ht!]
\centering
\caption{Track correction terms.}
\begin{tabular}{ c c c }
\hline
\hline
\textbf{Label} & \textbf{Coordinate} & \textbf{Calculation} \\
\hline
$\lambda_1$ & $\Delta$EL & $(X1 + X2) / 2$ \\
$\lambda_2$ & $\Delta$ X\_EL & $(Y1 + Y2) / 2$ \\
$\lambda_3$ & $\Delta$ X\_EL & $(X2 - X1)$ \\
\hline
\end{tabular}
\label{lamtable}
\end{table}

\begin{appendix}

\section{The pointing model equation}

Below are the full pointing model equations that utilize the terms that are further described in Tables~\ref{table1}, \ref{auxtable}, \ref{table3}, and \ref{lamtable} \citep{brandt2007}. Equations A1 and A2 provide the pointing models in cross-elevation and elevation, respectively. Servo hysteresis corrections are not included here. In the equations, EL represents the mount elevation angle and AZ represents the mount azimuth angle.

A brief note on sign conventions: in Tables~\ref{table1} and \ref{auxtable}, the signs of the coefficients match the conventions used by TPOINT; therefore, to reproduce the GBT pointing model in a TPOINT session, the values from the tables with their given signs can be plugged directly into a TPOINT model. However, due to differences between the coordinate conventions used by TPOINT and by the conventions adopted in the GBT control system, some of the coefficients in the model that is used by the antenna control software differ from the coefficients given in Tables~\ref{table1} and \ref{auxtable} by a sign flip. Plugging in the coefficients with their given signs from Tables~\ref{table1} and \ref{auxtable} to the equations below yields the formulation of the model as it is actually implemented by the GBT control system software; however, it should be noted that in the version of the model implemented in the software as of 2021, the track coefficient values (A12E, A13V, and A14A) have been updated from what is given in Table~\ref{auxtable} to account for the track's natural changes over time: 

\begin{align}
    \label{cross-el-model}
    \Delta X\_EL =\quad & CAL \nonumber \\
    &+ HSCE*cos(EL) \nonumber \\
    &+ HSSE*sin(EL) \nonumber \\
    &+ HSSASE*sin(AZ)*sin(EL) \nonumber \\
    &- HSCASE*cos(AZ)*sin(EL) \nonumber \\
    &- A13V*\lambda_2*sin(EL) \nonumber \\
    &+ A14A*\lambda_3*cos(EL) \nonumber \\
    &+ A2S*\tau_2 \nonumber \\
    &+ A3S*\tau_3 \nonumber \\
    &+ A4S*\tau_4 \nonumber \\
    &+ A10V*\tau_{10}*sin(EL) \nonumber \\
    &+ A11A*\tau_{11}cos(EL)
\end{align}
\begin{align}
    \label{el-model}
    \Delta EL =\quad &- IE \nonumber \\
    &- HESA*sin(AZ) \nonumber \\
    &+ HECA*cos(AZ) \nonumber \\
    &- HESE*sin(EL) \nonumber \\
    &- HECE*cos(EL) \nonumber \\
    &- A12E*\lambda_1 \nonumber \\
    &- A5E*\tau_5 \nonumber \\
    &- A6E*\tau_6 \nonumber \\
    &- A7E*\tau_7 \nonumber \\
    &- A8E*\tau_8 \nonumber \\
    &- A9E*\tau_9.
\end{align}

\newpage
\section{Figures}

\begin{figure}[h]
\centering
\includegraphics[width=0.5\linewidth]{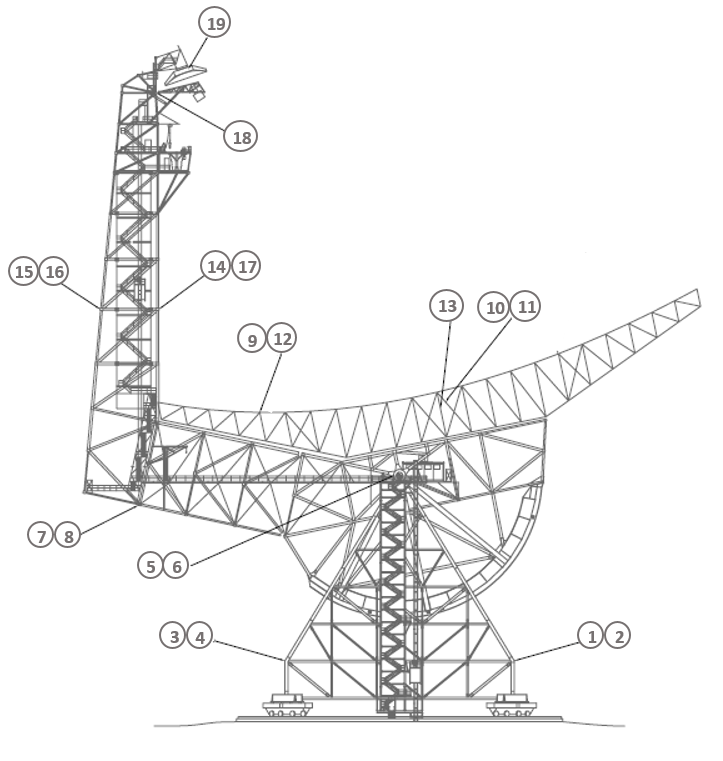}
\caption{Diagram showing GBT temperature sensor locations; numbers correspond to the numbers in Table~\ref{table2}.} 
\label{fig:temp1}
\end{figure}

\begin{figure}[h]
\centering
\includegraphics[width=0.5\linewidth]{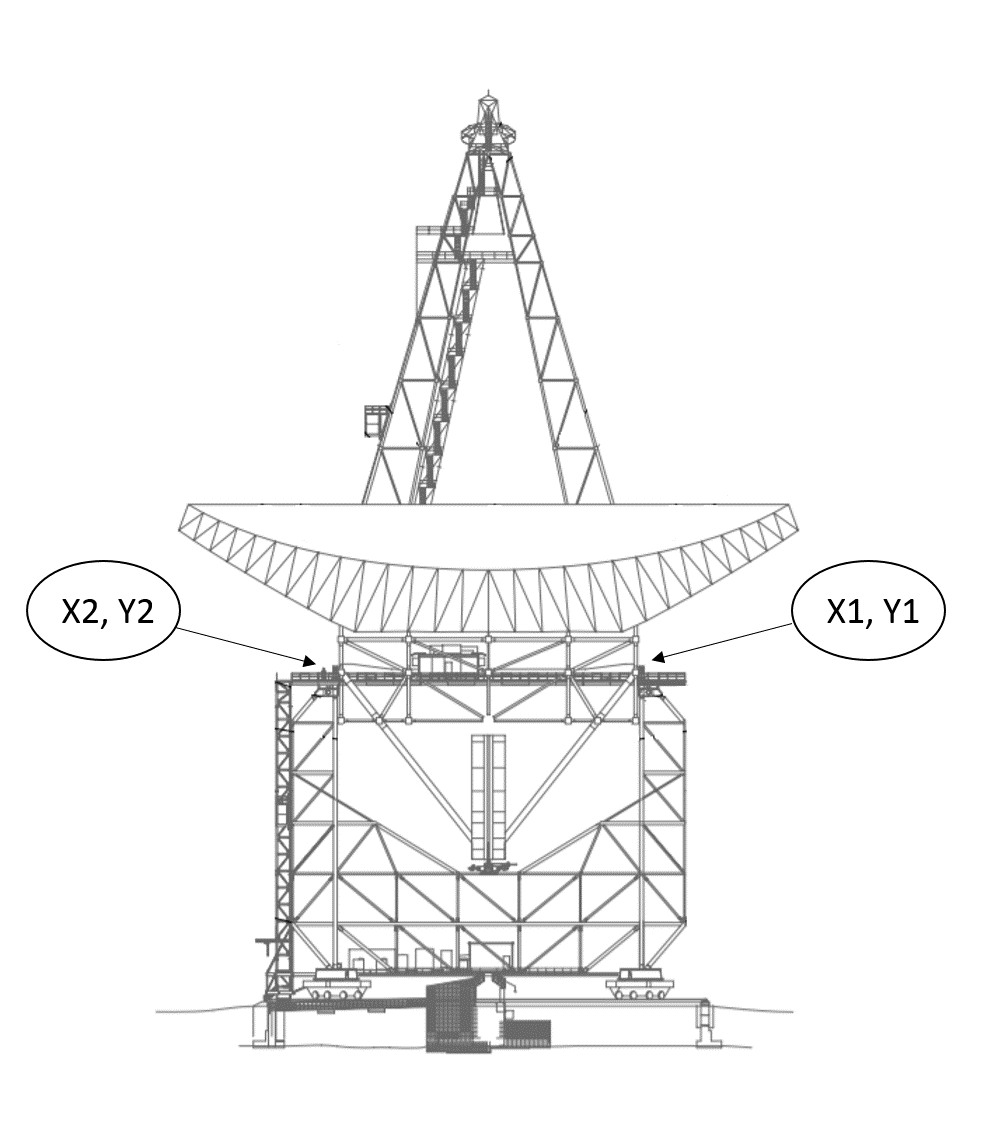}
\caption{Locations of the two inclinometers on the elevation bearing of the GBT.} 
\label{fig:incl}
\end{figure}

\begin{figure}[h]
\centering
\begin{subfigure}{0.8\textwidth}

\includegraphics[width=12cm]{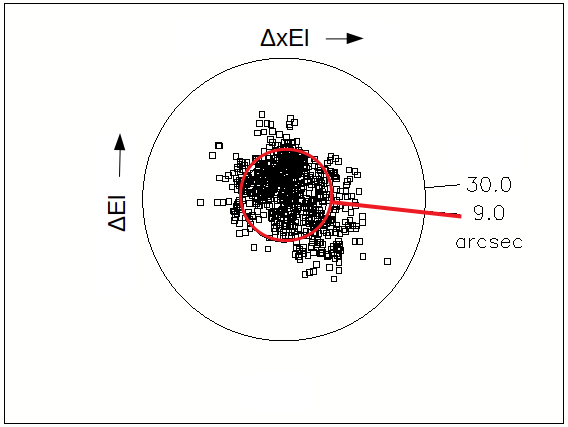}
\centering
\caption{Blind pointing}
\label{fig:bscat}
\end{subfigure}

\begin{subfigure}{0.8\textwidth}
\centering
\includegraphics[width=12cm]{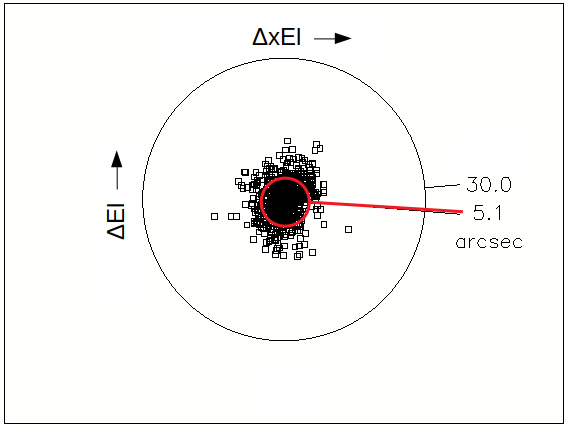}
\caption{All-sky relative pointing}
\label{fig:oscat}
\end{subfigure}

\caption{Pointing residual scatter plots. These diagrams show the error distribution with respect to the central beam of the GBT. Here, $\Delta El$ refers to the error in the elevation direction, and $\Delta xEl$ refers to the error in the cross-elevation direction. Blind pointing (a) refers to the pointing performance when the model is applied, but observers do not perform pointing calibrations at the beginning of a session. All-sky relative pointing (b) is the performance an observer would see if the model were applied and if a single pointing calibration were applied at the beginning of an observing session with sources distributed across the entire sky. The pointing measurements included in this plot were collected during nighttime pointing sessions taken between January 2017 and December 2018; wind speeds never exceeded 5 m/s for any of these measurements.}
\label{fig:scatter-plots}
\end{figure}

\begin{figure}[h]
\centering

\begin{subfigure}{0.7\textwidth}
\centering
\includegraphics[width=8cm]{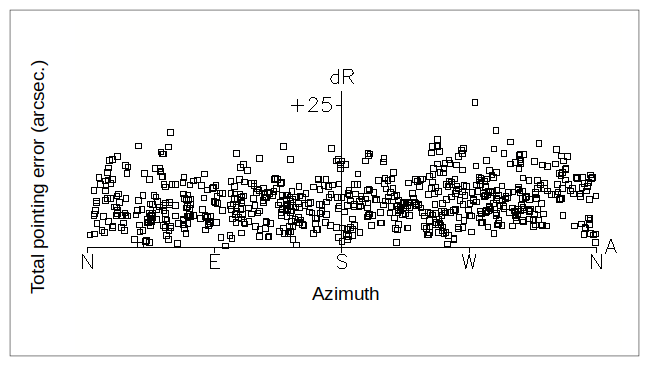}
\caption{Blind pointing}
\label{fig:baz2}
\end{subfigure}

\begin{subfigure}{0.7\textwidth}
\centering
\includegraphics[width=8cm]{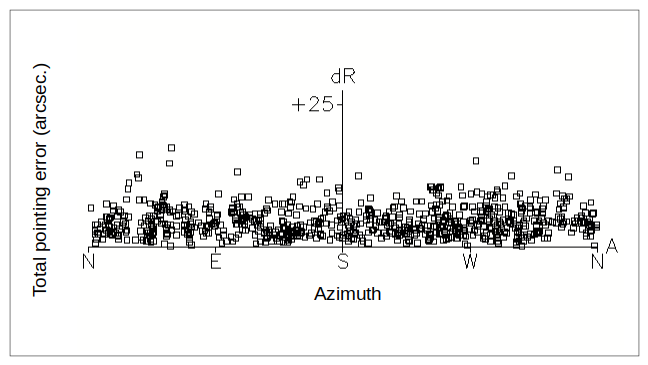}
\caption{All-sky relative pointing}
\label{fig:oaz2}
\end{subfigure}

\begin{subfigure}{0.7\textwidth}
\centering
\includegraphics[width=10cm]{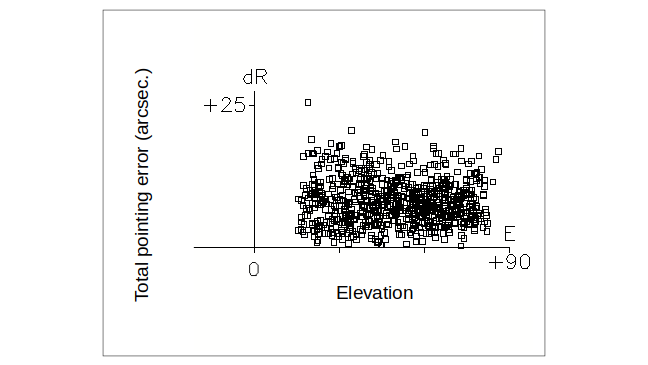}
\caption{Blind pointing}
\label{fig:bel}
\end{subfigure}

\begin{subfigure}{0.7\textwidth}
\centering
\includegraphics[width=10cm]{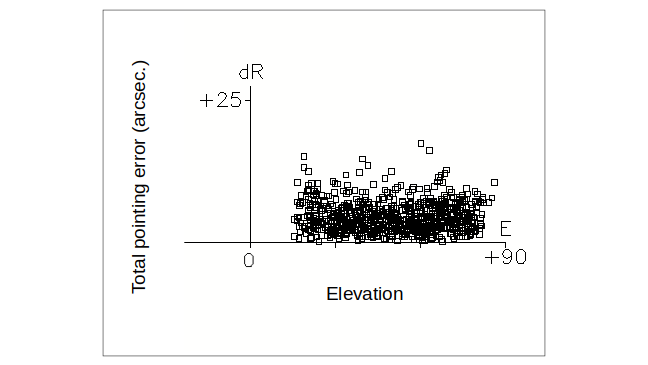}
\caption{All-sky relative pointing}
\label{fig:oel}
\end{subfigure}

\caption{Total pointing residual, dR (see Sect. 5 for a definition of dR), including both elevation and cross-elevation error, plotted versus elevation. Blind pointing refers to the pointing performance when the model is applied, but observers do not perform pointing calibrations at the beginning of a session. All-sky relative pointing is the performance an observer would see if the model were applied and if a single pointing calibration were applied at the beginning of an observing session with sources distributed across the sky, between a lower elevation limit of 15\degree \ and an upper limit of 87\degree.}
\label{fig:rve-plots}
\end{figure}

\end{appendix}

\end{document}